# Electrical control of spin coherence in ZnO

S. Ghosh[a)], D. W. Steuerman, B. Maertz, and D. D. Awschalom
*Center for Spintronics and Quantum Computation, University of California, Santa Barbara, CA 93106, USA*

K. Ohtani, Huaizhe Xu, and H. Ohno
*Laboratory for Nanoelectronics and Spintronics, Research Institute of Electrical Communications, Tohoku University, 2-1-1 Katahira, Sendai 980-8577, Japan*

Abstract

Electric field enhanced electron spin coherence is characterized using time-resolved Faraday rotation spectroscopy in *n*-type ZnO epilayers grown by molecular beam epitaxy. An in-plane *dc* electric field $E$ almost doubles the transverse spin lifetime at 20 K, without affecting the effective g-factor. This effect persists till high temperatures, but decreases with increasing carrier concentration. Comparisons of the variations in the spin lifetime, the carrier recombination lifetime and photoluminescence lifetimes indicate that the applied $E$ enhances the radiative recombination rate. All observed effects are independent of crystal directionality and are performed at low magnetic fields ($B < 0.2$ T).

---

[a)] Current address: School of Natural Sciences, University of California, Merced, CA 95344
Electronic address: sghosh@ucmerced.edu

A long-standing challenge toward the utilization of the spin of charge carriers in an electronic device has been the ability to manipulate these spins using electric fields in place of magnetic fields. Electric fields allow high-speed modulation and spatially localized application, both very difficult to achieve with magnetic fields. Considerable progress has been made in this regard in traditional II-VI and III-V semiconductors, such as ZnSe[1] and GaAs,[2] by exploiting the spin-orbit (SO) coupling of the valence band.

Oxide semiconductors have been receiving a lot of attention for their possible application as ferromagnetic semiconductors[3] with high transition temperatures that would allow their incorporation into operational data storage and processing devices. Research efforts in this area have particularly been focused on ZnO[4], which has led to interesting and promising experimental advances, such as, possible transition-metal doped ferromagnetism,[5,6] high quality nanofabrication and related quantum-confined phenomena [7,8] and room temperature spin coherence.[9] In this letter we explore the possibilities of electrical control of electron spin coherence in $n$-type ZnO, a material with practically negligible SO coupling.[10,11] While weak SO coupling allows for long electron spin lifetimes, with spin coherence in some cases persisting up to room temperature[9], it also limits the coupling of the spin degree of freedom to an applied electric field.

We study three 150 nm thick Ga-doped $n$-ZnO epilayer samples (samples A-C) with carrier density (mobility) of 2.1 x$10^{20}$ cm$^{-3}$ (46.5 Vcm$^2$/s), 2.6 x$10^{19}$ cm$^{-3}$ (36 Vcm$^2$/s) ), and 3.1x$10^{18}$ cm$^{-3}$ (46 Vcm$^2$/s), respectively. All the samples are grown on $a$-plane sapphire substrates with oxygen radical plasma assisted molecular beam epitaxy (MBE) at 750° C. The back of the sapphire substrates are mechanically polished to achieve optical transparency for measurements in transmission.

Figure 1a shows a schematic of the processed channels with reference to the spatial coordinates used in the measurements. X-ray diffraction data are used to identify the in-plane crystal axes orientations which allow etching channels along [100] and [110] directions. Each channel is 300 microns long (*l*) and 100 microns wide (*w*). They are contacted at the ends by Au (100-200 nm) / Ti (10-20 nm) deposits without annealing. We do not notice any deviation from Ohmic behavior in the temperature range used in our measurements (300 K-4 K). The samples are mounted on the variable temperature insert of a magneto-optical cryostat with the magnetic field *B* along the *y*-axis, in the plane of the samples and normal to the direction of optical spin injection, *z* (the Vogt geometry). The electric field *E* is generated by the application of a *dc* voltage along the length of the channels.

For the optical injection-detection of spin coherence, the time resolved Faraday rotation (FR) technique[12] is used. The output of a tunable, mode-locked Ti:sapphire laser is frequency-doubled to produce pulses with duration ~ 150 fs at a rate of 76 MHz. The wavelength is tuned to the absorption edge of each sample, and though it varies between samples and with temperature, it is typically in the range of 3.34-3.36 eV. The laser output is split into a circularly polarized pump and a linearly polarized probe beam (with powers 950 µW and 90 µW, respectively) which are focused to spot sizes of approximately 25 µm beam waist and spatially overlapped on the samples. The pump is incident normal to the sample surface and injects spin-polarized carriers with their magnetization along *z*. Faraday rotation of the probe ($\theta_F$), incident after a time delay $\Delta t$, is used to measure the remnant magnetization in the sample. The magnetic field causes the magnetization to precess about *B*, and $\theta_F$ obtained by continuously varying $\Delta t$ is an exponentially decreasing oscillatory trace (fig 1b) well characterized by $\theta_F = \theta_o \cos(g^* \mu_B B \Delta t/\hbar) \exp(-\Delta t/T_2^*)$, where $\mu_B$ is the Bohr magneton and $\hbar$ is Planck's constant. From this fit we can obtain the effective g-factor ($g^*$) of the carriers, the inhomogenous transverse spin lifetime, $T_2^*$ and the initial FR amplitude $\theta_o$. We measure a $g^*$ of 1.96, which agrees with previous measurements[9]. The inset to fig 1b shows $T_2^*$ in sample C at 20 K as a function of $B$. The decrease from 0.8 ns to 0.2 ns is not clearly understood, but has been seen in other systems and may be attributed to inhomogeneous dephasing[13]. Following this measurement, we limit our measurements to the low $B$ regime (0.2 T) for the rest of this letter. We have also not observed any dependence of any of the data on either crystal directionality or relative orientation of $E$ and $B$. All data shown in all samples in this letter is with $E \parallel [100]$ and $E \parallel B$.

Figure 1b compares the FR scans for $E = 0$ (top) and 4.9 mV/μm (bottom) in sample C at 20 K. The difference between the top and the bottom panels clearly show an enhancement in $T_2^*$. Fig 1c systematically follows the FR signal as a function of $\Delta t$ at 20 K as $E$ increases. $T_2^*$ derived from fits to these scans is shown in fig 2b (black) and is seen to almost double by $E = 5$ mV/μm. At higher $E$, local heating may be the cause of the leveling off of $T_2^*$. As we follow this effect to higher temperatures, the spin lifetime at $E = 0$ decreases, but the enhancement with $E$ persists, although less effective. At 100 K (red) $T_2^*$ at $E = 5$ mV/μm is about 75% more than that of the zero field value, while at 150 K (blue) the increase with a similar variation in $E$ is about 25%. The inset shows $T_2^*$ varying with T at $E = 0$ (red) and $E = 5$ mV/μm (black). The maximum enhancement of $T_2^*$ for all temperatures occurs in the vicinity of, but not exactly at, $E = 5$ mV/μm. We have verified that the effective g-factor $g^*$ remains unchanged as a function of electric field.

Fig 2b follows $T_2^*$ with $E$ in samples C and A at 20 K. $T_2^*$ at $E = 0$ is higher for sample A (higher carrier density) but practically no electrical enhancement is observed. Sample B (carrier density intermediate between samples A and C) shows results (not shown here) that establish the following trends: (a) $T_2^*$ (at $E = 0$) increases with carrier density in agreement with previous measurements[9] and (b) the electrical enhancement of $T_2^*$ is reduced as carrier density increases. We have not noticed any systematic trends, with either temperature or carrier concentration, in the value of $E$ at which $T_2^*$ is a maximum.

Fig. 2c and 2d show the variation of $\theta_o$ with $E$ for samples A and C at 20 K. While sample A shows a very small change in $\theta_o$ with $E$, a sharp decrease is noticed in sample C. A comparison of figs. 2b and 2d show that this decrease in the initial magnetization occurs over the same range of applied electric field over which $T_2^*$ is enhanced. Similar changes in the FR amplitude $\theta_o$ on application of an in-plane $E$ has been observed in other measurements[1,2] in samples with higher electron mobility and that variation was due to the drift of electrons over length scales larger than the spatial extent of the probe beam. In our samples this drift is on the scale of less than 1 µm for the largest applied $E$, well within the region investigated by the probe beam spot, and cannot be attributed to this simple explanation. A follow-up FR measurement with the pump and probe spots spatially separated is used to definitively verify this.

To understand the electrical enhancement of spin lifetimes and in particular, to see if there is any related effect on the carrier lifetimes, we perform time-resolved transmission (TRT) measurements. TRT is a pump-probe technique similar to time-resolved FR, except that it is not spin-resolved. The total intensity of the transmitted probe is mapped out as a function of $\Delta t$ and the trace obtained is a measure of the carrier lifetime. The pump and probe powers, energies and polarizations, measurement geometry, and magnetic field are same as those used for FR measurements for consistency. Fig. 3a shows TRT in sample C at 20 and 100 K as a function of $\Delta t$ at $E = 0$. Carrier lifetime measurements in bulk and epilayer samples of ZnO have been the subject of considerable investigation, and though results have varied depending on sample history and quality, typically at low temperatures (T < 160 K) two relaxation times have been observed[14,15]. Our transmission data also shows two decay times up to 150 K. Fitted to a bi-exponential, $A_o exp(-\Delta t/T_o) + A_1 exp(-\Delta t/T_1)$, at 20 K it yields a fast decay $T_1$ (90 ps) and a slow relaxation $T_o$ ( 0.5 ns). Both lifetimes decrease with increasing temperature and the inset follows the variation of the longer lifetime, $T_o$. By 150 K only one relaxation time is observed, on the scale of the faster decay.

We have also measured the spectrally-integrated time-resolved photoluminescence (PL) using a streak camera (resolution ~ 2 ps). In contrast to the TRT results, we observe a single exponential decay revealing PL lifetimes ($T_{PL}$) on the scale of the $T_1$ values (at 20 K $T_{PL}$ = 78 ps). While the pump-probe energies are tuned to the absorption edge for FR and TRT measurements, for PL we tune the excitation to 3.45 eV (360 nm). At 20 K, we observe a single, broad emission peak centered at 3.34 eV (371 nm). Measurements of PL lifetimes in ZnO samples have yielded a wide distribution of values depending strongly on sample purity and preparation techniques[16,17], but in general the experimental values of $T_{PL}$ have always been smaller than the theoretical radiative lifetime. This has been attributed to the presence of dominant non-radiative recombination pathways, since $1/T_{PL} = 1/T_R + 1/T_{NR}$, where $T_R$ and $T_{NR}$ are the radiative and non-radiative recombination times[18], respectively.

Fig. 3b plots the changes in transmission amplitude $A_o$ and lifetime $T_o$ with $E$ at 20 K. As the applied electric field is increased, $A_o$ decreases, similar to the variation in $\theta_o$ but contrary to the trend seen in the spin lifetime, $T_o$ decreases with increasing $E$. This decrease in $T_o$ persists till 100 K (fig. 3b, bottom) but the change in the $A_o$ is no longer as clearly delineated. Neither $T_1$ nor $T_{PL}$ show any change with electric field in any sample.

It would appear that the application of the in-plane electric field has a two-fold effect: it enhances the initial absorption causing a decrease in the transmission ($A_o$), while preferentially increasing the decay rate of the long-lived carriers without affecting the fast carrier relaxation rate. It is clear that the applied electric field enhances carrier relaxation along some pathway. This picture qualitatively explains our observations. The faster depopulation of the carriers with increasing $E$ results in both the reduced initial magnetization seen in the FR measurements, as well as the steady decrease in transmission amplitude. And while the decreased density of photogenerated carriers leads to smaller initial magnetization, it enhances the spin lifetimes by decreasing the decoherence caused by electron-electron interactions[19]. If the applied electric field was enhancing the non-radiative decay rate, $T_{PL}$ would not remain unchanged. In ZnO samples grown on sapphire substrates the PL is dominated by recombination of the bound exciton, and a simple calculation estimates the pure radiative lifetime of the bound exciton to be ~ 1 ns[20,21]. The measured $T_{PL}$ therefore clearly indicates a truncation of the radiative lifetime due to non-radiative processes. We could speculate that the applied field is in fact influencing the radiative decay rate, which is the longer relaxation time observed in TRT measurements. However, at this juncture, it is not possible to explain the exact mechanism of this effect without further investigation.

While a clear understanding remains yet to emerge, our observation has very important implications. The ability to manipulate spin coherence electrically is a very attractive proposition in semiconductors in general, as it bypasses the need for magnetic control. ZnO has generated a lot of interest as a prime candidate for exhibiting room temperature ferromagnetism. The results presented here, demonstrating an increase in the electron spin lifetimes by the application of a small, in-plane *dc* electric field, further adds to the attractiveness of this material as a candidate for spintronic applications.

The work was supported by the SRC FCRP, DMEA, the IT Program of Research Revolution 2002 (RR2002) from MEXT and Grant-in-Aids from JSPS/MEXT.

**Figure Captions:**

**Figure 1** (a) Schematic of processed structure used in measurements with the spatial coordinates for reference. (b) Time resolved Faraday rotation (FR) scans showing an in-plane electric field $E$ enhancing the spin lifetime by comparing scans at $E = 0$ (top) and $E = 4.9$ mV/μm (bottom). All measurements in this letter were done in the low magnetic field regime ($B = 0.2$ T) as the transverse spin lifetime ($T_2^*$) decreases with increasing magnetic field $B$ (inset). (c) Map of Faraday rotation ($\theta_F$) at $B = 0.2$ T as a function of electric field $E$ and delay time $\Delta t$ at 20 K. The scale of $\Delta t$ is magnified for better visualization of the enhancement effect. The amplitudes of all $\theta_F$ are normalized to their respective $\Delta t = 0$ values.

**Figure 2** (a) Spin lifetime obtained from fits to line-cuts of Fig 1c as a function of $E$ at different $T$ for sample C. Lines are guides to the eye. (Inset) $T_2^*$ at $E = 0$ (red) and 4.9 mV/μm (black) varying with $T$. (b) Spin lifetime changing with $E$ for samples A and C at $T = 20$ K and $B = 0.2$ T. Amplitude $\theta_o$ (normalized) of Faraday rotation at the same $T$ and $B$ for (c) sample A and (d) sample C, varying with $E$.

**Figure 3** (a) Time resolved transmission for sample C at $T = 20$ and 100 K at $E = 0$. (Inset) The slow relaxation time ($T_o$) obtained from bi-exponential fits to traces in main figure as a function of $T$. Transmission amplitude $A_o$ and relaxation time $T_o$ varying with $E$ at 20 K (b) and 100 K (c).

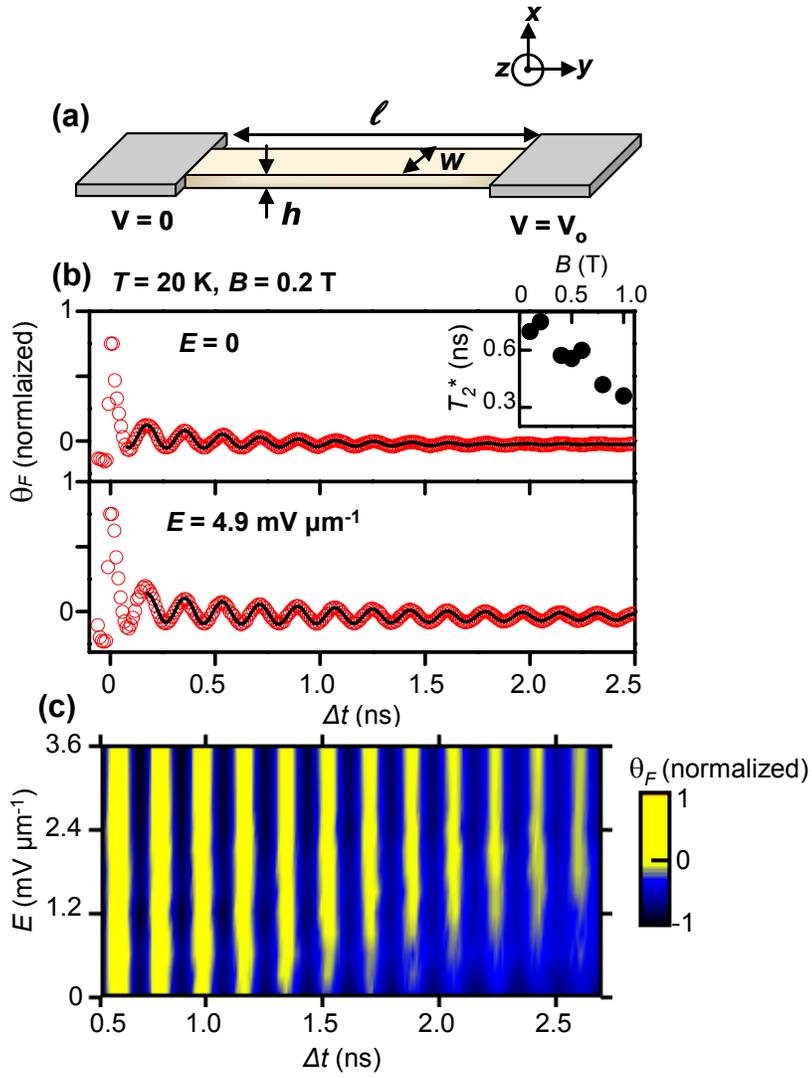

**Figure 1**
**S. Ghosh, *et. al.***

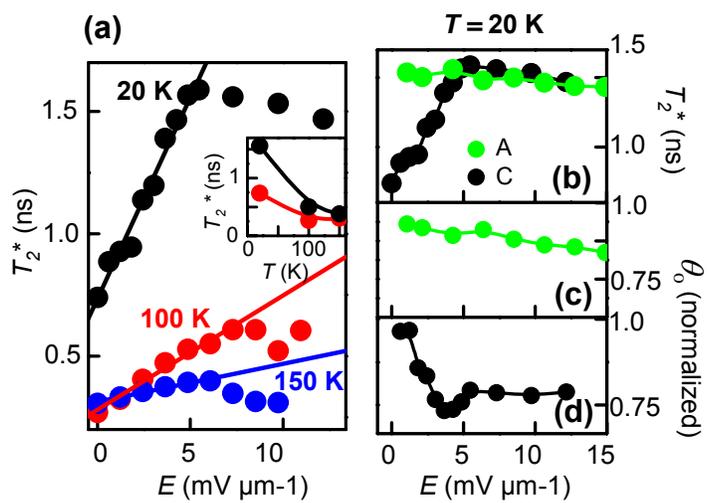

**Figure 2**
S. Ghosh, *et. al.*

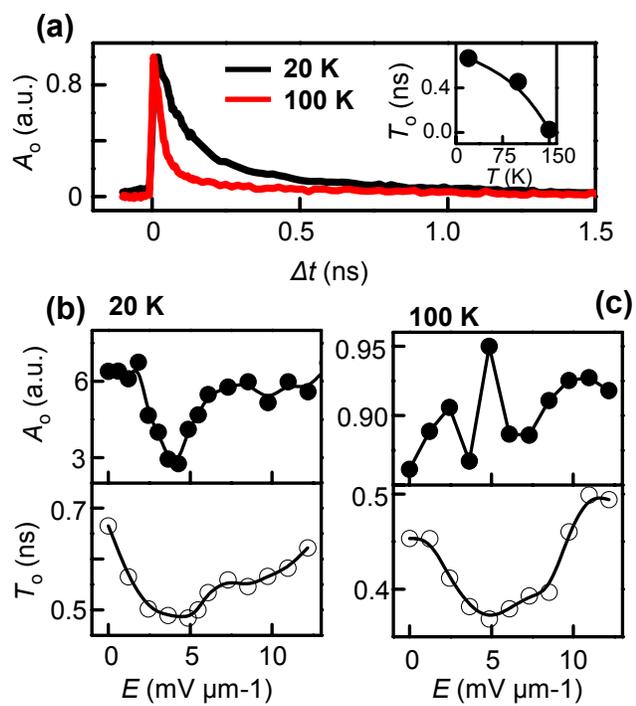

**Figure 3**
**S. Ghosh, *et. al.***